# Higher-Order Interactions in Brain Connectomics: Implicit versus Explicit Modeling Approaches

Mohamma Reza Salehi[1], Ali BashirGonbadi[2], Hamid Soltanian-Zadeh[3]

1. Ali Bashirgonbadi is with the Control and Intelligence Processing Center of Excellence (CIPCE), School of Electrical and Computer Engineering, College of Engineering, University of Tehran, Tehran 14399, Iran (e-mail: a_bashirgonbadi@ut.ac.ir).
2. Mohamad Reza Salehi is with the Control and Intelligence Processing Center of Excellence (CIPCE), School of Electrical and Computer Engineering, College of Engineering, University of Tehran, Tehran 14399, Iran and School of Cognitive Sciences, Institute for Research in Fundamental Sciences (IPM), Tehran, Iran (e-mail: mohamadrezasalehi@ut.ac.ir).
3. Corresponding author: Hamid Soltanian-Zadeh is with the Control and Intelligence Processing Center of Excellence (CIPCE), School of Electrical and Computer Engineering, College of Engineering, University of Tehran, Tehran 14399, Iran; School of Cognitive Sciences, Institute for Research in Fundamental Sciences (IPM), Tehran, Iran and Image Analysis Laboratory, Departments of Radiology and Research Administration, Henry Ford Health System, Detroit, MI 48202, USA (e-mail: hszadeh@ut.ac.ir).

## Abstract
The human brain is a complex system defined by multi-way, higher-order interactions invisible to traditional pairwise network models. Although a diverse array of analytical methods has been developed to address this shortcoming, the field remains fragmented, lacking a unifying conceptual framework that integrates and organizes the rapidly expanding methodological landscape of higher-order brain connectivity. This review provides a synthesis of the methodologies for studying higher-order brain connectivity. We propose a fundamental distinction between implicit paradigms, which quantify the statistical strength of group interactions, and explicit paradigms, which construct higher-order structural representations like hypergraphs and topological data analysis. We trace the evolution of each approach, from early Correlation-of-Correlations and information-theoretic concepts of synergy/redundancy, to the edge-centric paradigm and advanced topological methods. Through a critical analysis of conceptual, statistical, and computational challenges, we argue that the future of the field lies not in a single best method, but in a principled integration of these complementary approaches. This manuscript aims to provide a unified map and a critical perspective to guide researchers toward a robust and insightful understanding of the brain's complex, multi-level architecture.

# 1. Introduction

The history of neuroscience is, in many ways, a story of successful reductionism. For decades, the mapping of specific cortical areas by pioneers like Broca and Wernicke, the dominant paradigm focused on decomposing the brain into its constituent parts, individual neurons, columns, and localized regions, to understand their specialized functions [1]. This approach has been immensely fruitful, providing the foundational knowledge upon which modern neuroscience is built. However, as our understanding of complex cognitive phenomena like consciousness, learning, and decision-making has grown, the limitations of a purely localizationist view have become increasingly apparent. It is now clear that brain activity is not an isolated phenomenon but the result of dynamic, integrated interactions among distributed neural populations [1], [2], [3], [4]. This realization catalyzed the rise of network neuroscience, a powerful framework that models the brain as a graph of nodes and edges, providing a comprehensive system-level picture of its functional organization [2].

While network neuroscience represented a major step away from simple localizationism, much of its early development remained rooted in a form of relational reductionism, focusing primarily on pairwise analyses of brain connectivity. This dyadic paradigm, while computationally tractable and highly informative, inherently simplifies the complex, multi-way interactions that define a truly integrated system [5], [6], [7].

The brain, as a complex system, exhibits emergent properties where the behavior of the whole cannot be fully predicted by the sum of its pairwise parts [8]. A growing body of evidence, from physiological studies of neuromodulation to computational models, suggests that the interaction between two regions is often fundamentally altered by the concurrent activity of a third, a principle that pairwise models cannot capture [9], [10]. This has motivated a new concept toward higher-order frameworks that implicitly or explicitly model interactions among more than two brain regions, extending the traditional pairwise view rather than replacing it.

Remarkable conceptual and mathematical diversity characterize this new frontier. The very definition of a higher-order interaction varies across subfields, encompassing a wide spectrum of ideas from second-order correlations to more abstract concepts like information-theoretic synergy and edge-centric perspectives. Despite this exciting growth, the literature remains fragmented.

The primary aim of this review is to provide a comprehensive and conceptually grounded roadmap to the rapidly growing field of higher-order interaction analysis in neuroscience. To bring coherence to this diverse and often fragmented landscape, we organize existing methodologies into two overarching paradigms defined by the nature of their analytical output: Implicit Methods and Explicit Methods, with further subdivisions capturing key conceptual and methodological distinctions (Figure 1). The first category, Implicit Methods, encompasses approaches that estimate the strength and character of group interactions without explicitly redefining the network structure. These methods typically yield a numerical metric or an enriched pairwise representation and include frameworks based on Correlation-of-Correlations (CoC), multivariate statistical dependencies, and information-theoretic measures. The second category, Explicit Methods, comprises approaches that construct new mathematical objects, such as hypergraphs and Simplicial Complexes, that directly encode multi-way relationships among brain regions. In addition, we discuss hybrid approaches, as well as recent developments extending higher-order analyses to structural brain networks. Figure 1 provides a visual taxonomy of the frameworks covered in this review, serving as a guide for the detailed methodological survey that follows.

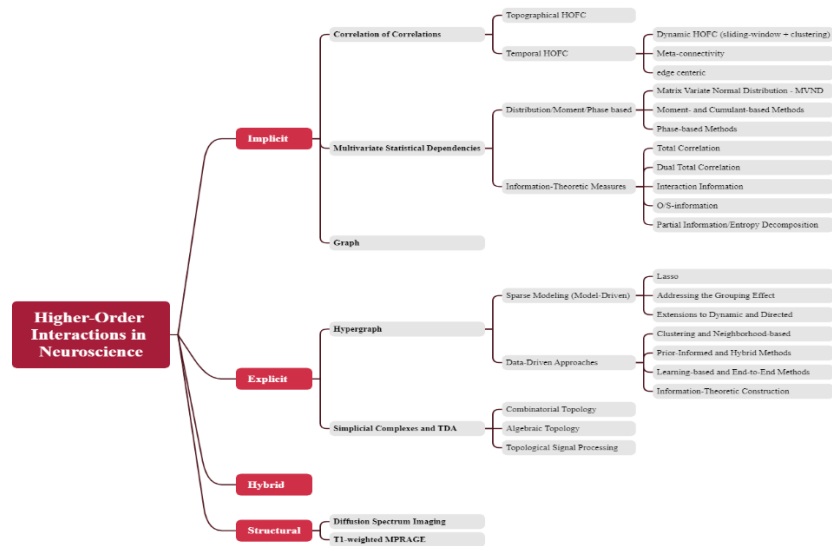

Figure 1. A Conceptual Taxonomy of Higher-Order Interaction Analysis in Neuroscience.

This methodological survey forms the core of our review. To further aid navigation, a comprehensive mind map illustrating the structure and key concepts of this methodological review is provided as a supplementary file (Supplementary Figure S1). Following this, we will transition from description to critique, presenting a dedicated analysis of the grand challenges and limitations that confront the field, spanning conceptual, statistical, and computational hurdles. The review will then conclude with a discussion of key comparative insights and a forward-looking perspective on the future directions that promise to shape this exciting frontier. We hope this structured overview will serve as an essential reference for any researcher aiming to navigate and contribute to the complex, entangled, and truly higher-order nature of the human brain.

## 2. Methods

The pursuit of understanding higher-order interactions has led to a vast methodological landscape. To navigate this diverse terrain, we organize the existing approaches into two principal paradigms defined by the nature of their analytical output: Implicit methods, which quantify the statistical signature of group interactions, and Explicit methods, which construct new mathematical objects to represent them directly.

Implicit methods aim to quantify the statistical signature of a multi-way interaction, typically yielding a numerical measure, such as a scalar value representing the strength of a group interaction or an enriched pairwise network that embeds higher-order information. The analytical focus in this paradigm is on the robust statistical estimation and interpretation of the dependency measure itself. In contrast, explicit methods are fundamentally concerned with constructing new, tangible mathematical objects, namely hypergraphs and Simplicial Complexes, that directly represent group interactions as their primary output. Here, the analytical challenge shifts from estimation to the construction and subsequent topological or structural characterization of these objects. This following section will systematically survey the key methodologies within each of these two major paradigms, beginning with the diverse family of implicit approaches.

### 2.1: Implicit Methods

As outlined, the implicit paradigm seeks to quantify the statistical signature of higher-order interactions, typically yielding a numerical measure or an enriched pairwise network. The analytical challenge in this paradigm lies in the robust statistical estimation and interpretation of the dependency measure itself. Our survey of these methods begins with the intuitive CoC framework, followed by approaches based on multivariate statistics and information theory.

The implicit paradigm encompasses a rich and diverse family of methodologies, each offering a unique lens through which to view higher-order dependencies. Our survey begins with the intuitive and widely-used CoC framework, which extends the concept of pairwise similarity to a second order. We then broaden our scope to a diverse family of multivariate statistical dependencies, including methods based on statistical distributions, moments, and phase dynamics. Following this, we delve into the powerful, model-free framework of information-theoretic measures, which provide a principled way to decompose complex interactions into fundamental components like redundancy and synergy. Finally, we will touch upon dynamic graph theory and statistical physics. This comprehensive survey will illuminate the diverse strategies available for quantifying the intricate, beyond-pairwise relationships that define brain function.

### 2.1.1: Correlation of Correlations

At the foundational level of conceptualizing higher-order interactions in the brain, CoC approaches have emerged as one of the earliest yet most influential frameworks [11], [12], [13], [14], [15], [16]. In this methodology, the analysis begins with estimating either static or dynamic correlations across brain regions, commonly referred to in the literature as lower-order functional connectivity (LOFC). Subsequently, by quantifying the similarity or correlation among these LOFC patterns, a secondary network is constructed, representing HOFC [11], [12]. This procedure essentially reflects a second-order correlation, as it builds upon the interrelationships between first-order connectivity patterns. Depending on whether the second-order correlations are applied to the spatial dimension [12] or to the temporal profiles [11] of the first-order networks, HOFC can be classified into two primary categories: topological and temporal HOFC [11], [12].

*2.1.1.1: **Topological higher-order functional connectivity***

Among the first and most influential frameworks of the CoC principle is topological HOFC. Conceptually, topological HOFC can be understood through an analogy to social networks: it measures the extent to which two brain regions share a similar community of connections or circle of friends. This method is designed to uncover higher-level associations by evaluating the similarity of their LOFC profiles (illustrated in Figure 2A). For each brain region, a connectivity profile is constructed, representing its Pearson correlation pattern with all other regions in the brain (equivalent to a row or column in the LOFC matrix). The similarity between any two such profiles is then computed, typically via Pearson correlation, to form the topological HOFC matrix. Unlike direct time-series correlations, topological HOFC captures second-order relationships by quantifying the degree to which two regions share similar connectivity patterns with the rest of the brain [12]. Beyond providing complementary information and demonstrating distinct predictive value relative to lower-order networks [15], [17], [18], [19], [20], topological HOFC enhances the characterization of complex modular architectures [12], [21], [22] and exhibits greater sensitivity to subtle shifts in connectivity patterns [12], [15], [20], [23]. A schematic overview of this extraction process is illustrated in Figure 2B.

Building upon the foundational topological HOFC framework, several studies have explored its deep extension to capture deeper and more complex inter-level relationships. One direct extension involves deriving correlations of ever-increasing order through iterative applications of the CoC procedure. In this framework, the connectivity matrix produced at one order serves as the input for computing the next, allowing for a recursive extension to orders n > 2. Such an approach enables the representation of more complex, hierarchical, and multilayered interactions within brain network organization [19], [24]. A conceptually related development is associated HOFC (aHOFC), which was proposed to measure the similarity between a brain region's own LOFC profile and its corresponding topological HOFC profile. By quantifying this cross-level interaction, aHOFC provides a unique view of the brain's organizational hierarchy and has proven effective in diagnostic applications [17]. A schematic overview of the computation of third-order FC and aHOFC is presented in Figures 2C and 2D, respectively.

Another major trajectory in the evolution of the topological HOFC framework has been its superficial extension, broadening its scope from individual brain regions to larger organizational units. One such formulation extends the concept to large-scale functional networks, where the connectivity profile of an entire network is derived by averaging across its constituent nodes. The correlations among these network-level profiles then yield the inter-network HOFC (IN-HOFC), providing insights into the coordination between large-scale brain systems [25]. A conceptually related approach has also been advanced at the inter-subject level, wherein the similarity of LOFC profiles across different individuals is assessed for a given brain region. Such inter-subject analyses offer a complementary perspective, enabling the detection of both shared and group-specific network architectures and thereby contributing to the characterization of common versus divergent patterns of functional organization across populations [26], [27].

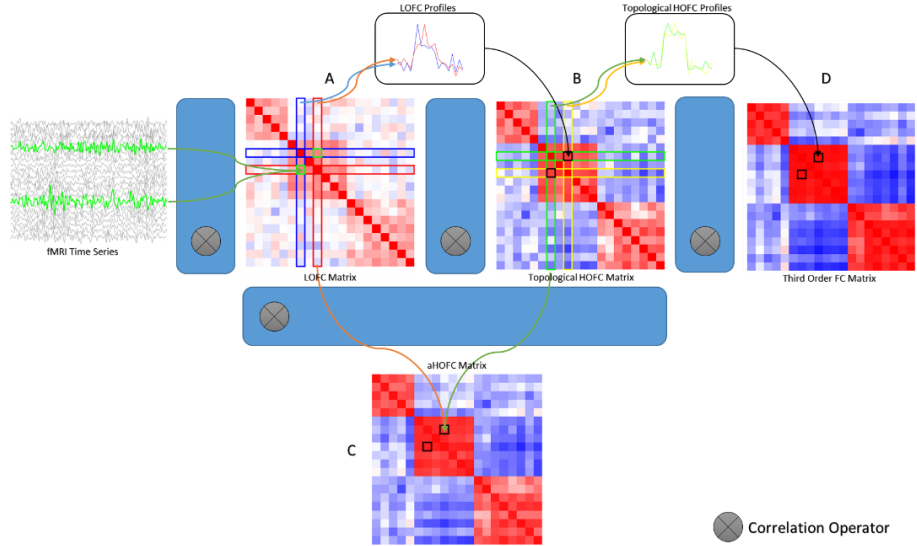

*Figure 2 Schematic illustration of Topological Higher-Order Functional Connectivity (tHOFC) and its extensions. (A) Low-Order Functional Connectivity (LOFC): Pearson correlation between the fMRI time series of any two ROIs (e.g., green traces). Each entry in this matrix represents a direct, pairwise functional connection. (B) Topological HOFC (Second-Order): The LOFC matrix serves as the input for the next level. The connectivity profile of a given ROI (e.g., the red and blue rows/columns in the LOFC matrix) represents its pattern of connections to all other brain regions. The correlation between two such profiles (e.g., red and blue traces in "LOFC Profiles") is then calculated. This second-order correlation value populates the corresponding entry in the Topological HOFC (tHOFC) matrix, capturing the similarity of connectivity patterns between the two ROIs. (C) Associated HOFC (aHOFC): This metric quantifies the cross-level interaction for a single ROI. It is computed as the correlation between that ROI's own LOFC profile (e.g., a row from the LOFC matrix) and its tHOFC profile (e.g., the corresponding row from the tHOFC matrix). The resulting aHOFC matrix captures the relationship between first- and second-order connectivity architectures. (D) Third-Order FC: The process can be iterated. By treating the tHOFC matrix as a new input, the correlation between its profiles (e.g., green and yellow traces in "Topological HOFC Profiles") is computed to generate a Third-Order FC matrix. This reveals even higher levels of network organization. The crossed circle symbol represents the correlation operator applied at each stage.*

To capture the brain's non-stationary nature, the topological HOFC framework has been extended into the temporal and spectral domains. The most common approach, dynamic topological HOFC, involves dividing time series into overlapping windows, estimating LOFC within each window, and then transforming these into a sequence of time-varying topological HOFC matrices [17], [28], [29], [30], [31]. This strategy enables the temporal tracking of higher-order connectivity patterns and has been further adapted by applying it across different frequency bands to generate time-frequency-specific topological HOFC matrices, particularly in EEG studies [18], [32], [33], [34]. To extract informative features from these dynamic higher-order networks, novel analytic methods have been proposed, such as the central-moment approach, which characterizes temporal correlation dynamics through higher-order statistical moments [28], [35].

Recent research has increasingly focused on enhancing the topological HOFC framework through methodological innovations at various stages of the pipeline, often involving the fusion of multiple, complementary information sources. Innovations in the pre-computation stage aim to enrich the input LOFC matrix. For example, some studies construct a more comprehensive LOFC by fusing spatio-temporal features before applying the topological HOFC procedure [36], while others employ Correlation-Preserving Embedding (COPE) to learn a richer, topology-aware representation of brain regions prior to computing profile similarity [37], [38]. Another innovative direction involves replacing the standard correlation-based FC with effective connectivity (EC) as the low-order network [39]. Methodological creativity is also evident in the use of novel data sources, such as incorporating White Matter (WM) BOLD fluctuations to construct cross-tissue topological HOFC networks [40], and in post-processing refinements like low-rank matrix factorization to reduce noise and enhance the modular organization of the final topological HOFC matrix [21]. An evolution of the framework involves embedding the construction of hierarchical-order networks directly within an end-to-end, task-oriented learning model. For instance, Liu et al. (2025) proposed a framework where hierarchical-order connectivity is learned iteratively using an attention mechanism, ensuring that the resulting high-order relationships are explicitly optimized for a specific predictive goal, such as amyloid-β deposition prediction [41].

One of the recent trends is the integration of topological HOFC as one of several complementary views within multi-view fusion frameworks. The underlying premise is that different FCN construction methods (e.g., Pearson correlation, sparse representation, mutual information, and topological HOFC) capture distinct aspects of neural interactions. Rather than seeking a single best representation, these studies construct a portfolio of FCNs and employ advanced fusion strategies, such as deep feature-level fusion [42], adaptively learn the fusion weights [43], joint embedding and tensor decomposition [44], or explicit separation of shared versus unique network information [45]. This multi-view philosophy has been powerfully integrated with deep learning. Graph Neural Networks (GNNs), for instance, have been used to process multi-level graphs composed of both low-order and higher-order networks, often within a joint learning framework that allows for information exchange between the different network levels [46], [47], [48], [49], [50]. Similarly, attention-based models are used to dynamically weigh the importance of features derived from different network views [51]. Overall, this multi-view fusion paradigm is predicated on the principle that a more holistic and robust representation of brain functional architecture can be achieved by integrating these complementary information sources. Empirical evidence consistently supports this premise, showing that fusion-based models significantly enhance the performance of downstream analyses, particularly in the classification of neurological and psychiatric disorders [44], [45], [48], [51].

In summary, the topological HOFC framework has evolved substantially from its original formulation based on LOFC profile similarity into a diverse family of approaches. Through deep, superficial, temporal, and methodological extensions, it has been adapted to capture interactions across multiple scales, levels, and data modalities. These developments have been primarily driven by the goal of enhancing sensitivity to the brain's complex network architecture. Importantly, the application of topological HOFC and its variants has yielded significant clinical and neurobiological insights. It has consistently improved the classification and characterization of a wide range of conditions, including Alzheimer's disease and its prodromal stages [12], [17], [29], Parkinson's disease [15], presbycusis [20], schizophrenia [22], and autism spectrum disorder [19]. Furthermore, it has been used to predict clinical outcomes, such as post-stroke somatosensory function [23], and to provide neurobiological insights into complex theories like the integration-segregation imbalance in schizophrenia [22].

*2.1.1.2: Temporal higher-order functional connectivity*
In parallel with methods that assess the spatial similarity of connectivity patterns (topological HOFC) [12], a distinct and highly influential family of approaches quantifies higher-order interactions by examining the temporal similarity of these patterns [11], [52], [53]. The core principle uniting these methods is the concept of a second-order temporal correlation, and they measure how the strengths of different functional connections fluctuate together over time. The general computational workflow for this paradigm, which we term Temporal HOFC, is illustrated in Figure 3. A useful analogy is to consider temporal HOFC as measuring conversational synchrony in a dynamic social gathering. While individual conversations (lower-order connections) fluctuate in intensity, this family of approaches identifies which pairs of conversations tend to strengthen or weaken in unison. This conceptual framework has evolved along several parallel, yet interconnected, research trajectories, often appearing under different terminologies. To provide a unified perspective, we introduce the umbrella term Temporal Higher-Order Functional Connectivity (Temporal HOFC) to encompass these methods. Key paradigms within this family include the dynamic HOFC (dHOFC) approaches [11], the concept of meta-connectivity [54], and a high-resolution edge-centric paradigm [52]. While a related concept, the Matrix Variate Normal Distribution (MVND) framework, also captures the covariance of connections, it does so through a unified statistical estimation process and will be discussed in a subsequent section [55]. In the following, we will systematically survey the evolution and application of these temporal correlation-based methods.

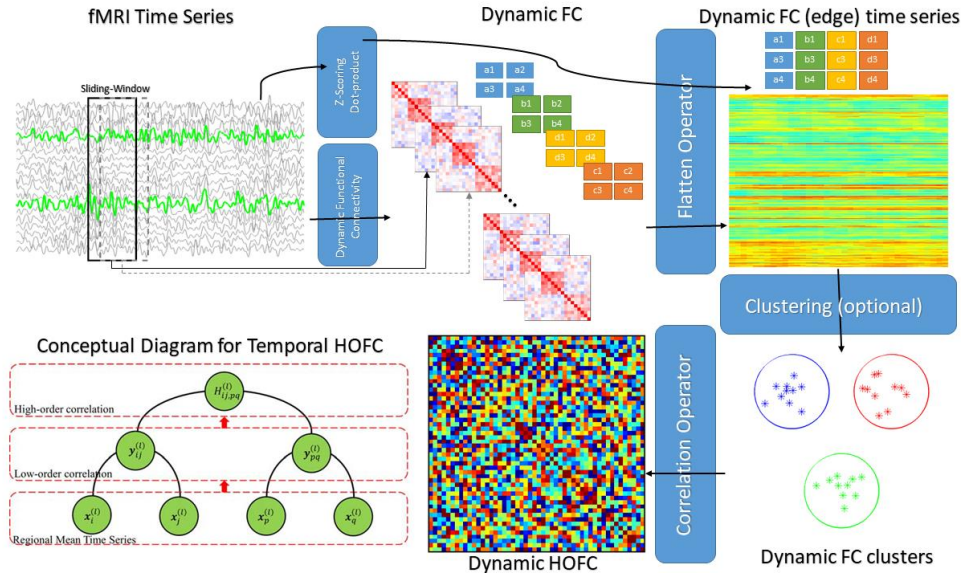

*Figure 3. Computational pipelines for Temporal Higher-Order Functional Connectivity (Temporal HOFC). In both cases, the result is a matrix of Dynamic FC (edge) time series, where each row represents the temporal fluctuation of a single pairwise connection (edge). A Flatten Operator conceptually transforms the sequence of matrices into this edge-by-time representation. Subsequently, a Correlation Operator computes the pairwise correlation between these edge time series, yielding the final Dynamic HOFC matrix. This matrix captures the second-order temporal relationships, i.e., the co-fluctuation patterns between different brain connections. As an optional intermediate step, the edge time series can be grouped using a Clustering algorithm to identify Dynamic FC clusters or "meta-modules" before computing the final correlations. The Conceptual Diagram (bottom-left [11]) summarizes this hierarchical process, showing how high-order correlations are built upon low-order correlations, which themselves derive from the regional mean time series.*

Dynamic DOFC: The first introduction of this family was dHOFC. The methodology begins by employing a sliding-window analysis on fMRI time series to generate a sequence of time-varying, low-order FC (dFC) matrices. From this sequence, the temporal trajectory of each individual connection (i.e., each edge in the low-order network) is extracted, forming a correlation time series. The final dHOFC network is then constructed by computing the Pearson correlation between these correlation time series. Conceptually, this process transforms the original pairwise edges of the LOFC into the nodes of the new dHOFC network, where the edges now represent the temporal synchrony of their fluctuations. Ultimately, dHOFC provides a higher-order representation of the brain's functional organization by mapping the underlying architecture of its temporal dynamics [11].

A direct consequence of this formulation is the curse of dimensionality; calculating correlations between all possible pairs of edges ($N^2$) results in a feature space on the order of $N^4$, which is computationally prohibitive and statistically challenging. To address this, a common and effective solution is the application of clustering algorithms [11], [56], [57], [58], [59]. In this modified pipeline, the numerous correlation time series are typically pooled across all subjects to create a common feature space and then grouped into a smaller number of clusters using methods such as hierarchical [11], [58], [59], [60] or k-means [16], [61] clustering. Crucially, the number of clusters is often treated as a hyper parameter, with researchers searching across a range of values to identify the optimal setting that maximizes performance on a downstream task, such as clinical classification [11], [58], [59]. The final dHOFC network is then computed based on the correlations between the mean time series of these clusters, significantly reducing the dimensionality while retaining the core dynamic patterns [11], [56], [57].

Building on this core concept, the dHOFC framework has evolved along several distinct methodological lines. One major development has been the introduction of hierarchical structures, where multi-layer dHOFC networks are constructed by systematically varying the number of clusters across layers, allowing for a multi-scale analysis of dynamic organization [14]. Another prominent trajectory involves fusing dHOFC-derived features with those from other connectivity measures, such as topological HOFC [60], [62], static or dynamic low-order FC, to create a more comprehensive and powerful representation of brain function for classification tasks [63], [64], [65], [66]. Also, the field has moved toward integrating dHOFC with Graph Neural Networks (GNNs), which can automatically learn complex topological and nodal features from these higher-order graphs, representing a significant advance over manual feature engineering [67].

Other methodological refinements have sought to improve specific stages of the dHOFC pipeline. These include the development of adaptive dFC estimation methods that move beyond the limitations of the rigid sliding window [68] and the use of central moments to derive time-invariant statistical summaries of the clustered correlation time series, thereby creating features that are robust to temporal permutations [69]. Furthermore, the dHOFC framework has proven versatile, having been successfully adapted for other imaging modalities such as EEG [70] and combined with structural methods like the Minimum Spanning Tree (MST) for feature refinement [71].

Meta-Connectivity: Closely related to the dHOFC framework is the concept of meta-connectivity (MC), a term often used to describe the correlation of dynamic functional connectivity fluctuations [54], [72]. hile methodologically similar to dHOFC in its reliance on sliding-window estimates of dFC, meta-connectivity emphasizes a shift in perspective towards treating the dynamic links themselves as the primary interacting units of a complex system. This framework has been particularly influential in characterizing the temporal organization of network dynamics beyond simple state-based models. For instance, MC analysis has revealed that brain network dynamics are not random but are organized into meta-modules, groups of functional links that co-fluctuate over time, and are coordinated by specific "meta-hubs" that act as control centers for network reconfiguration [54]. Recent applications have further leveraged this concept, for example, to identify distinct state-switching dynamics disrupted in Alzheimer's disease [53]. or to develop lightweight, attention-based Graph Neural Network (GNN) models for classifying autism spectrum disorder, where MC matrices serve as rich, higher-order edge features [73].

The Edge-Centric Paradigm: A particularly fruitful and distinct line of inquiry within the temporal HOFC family has been the edge-centric approach. This paradigm reformulates the problem by treating the dynamic edges themselves as the fundamental units of analysis, moving beyond node-centric views to directly model the interactions between pairwise connections [52], [74]. The key innovation that propelled the edge-centric paradigm was the introduction of a parameter-free, high-temporal-resolution estimate of dynamic connectivity, addressing the limitations of traditional sliding-window techniques such as temporal blurring.

The cornerstone of this approach is the concept of the Edge Time Series (ETS), first proposed by Zamani Esfahlani et al. to uncover the event-driven nature of static FC [75], and formalized into a higher-order framework by Faskowitz et al. [52]. An ETS is generated for each pair of brain regions by computing the element-wise product of their respective z-scored BOLD time series [75]. The resulting value at each time point captures the instantaneous co-fluctuation of that specific edge, providing a frame-by-frame account of dynamic connectivity without the blurring effects inherent to sliding-window methods. The subsequent correlation of these ETS between different pairs of connections yields the Edge Functional Connectivity (eFC) matrix, a higher-order network where nodes represent the original connections and the new edges represent the similarity of their dynamic profiles [52]. This framework has been foundational, with its principles and applications extensively reviewed and expanded upon in subsequent work [76], [77].

The development of ETS has catalyzed a rich and expanding field of research focused on the brain's fine-scale temporal dynamics, a literature too vast to comprehensively survey here. However, its foundational findings are critical for understanding the motivation behind eFC. A key discovery from the direct analysis of ETS was the identification of high-amplitude events: brief, intermittent moments of strong, brain-wide co-fluctuation that were found to be the primary drivers of the overall static functional connectivity structure [75]. This event-based perspective has proven highly insightful, and as a few illustrative examples, studies have used the spatial patterns of these events to demonstrate their link to the brain's underlying structural modularity [78]; others have used these specific temporal moments to uncover robust individual fingerprints of brain connectivity [79]; and some have leveraged the principle of instantaneous co-fluctuation to characterize the connectome profiles of discrete brain states, such as dynamic co-activation patterns (CAPs) [80]. While the direct analysis of ETS dynamics continues to be a fruitful area of research in its own right, our focus in this review is on its role as a precursor to a truly higher-order measure: the correlation between these dynamic edge profiles, known as eFC [52], forming a truly higher-order network.

The analysis of this network, most notably through the discovery of edge communities, has yielded profound insights into brain architecture. By applying clustering algorithms to the eFC matrix, researchers can identify groups of connections that exhibit similar co-fluctuation patterns over time. A key and powerful consequence of this approach is the natural emergence of overlapping modules at the node level; since a single brain region can be an endpoint for multiple edges belonging to different communities, this framework inherently reflects the multifunctional nature of brain regions [52]. This technique has been instrumental in studying the complex organization of cortical [52] or cortico-subcortical interactions [81], uncovering the hierarchical and self-similar nature of overlapping modules [82], and understanding their development across childhood and adolescence [83]. Furthermore, the concept of edge-community entropy has been introduced as a novel metric to quantify the degree of a node's participation in these multiple communities, providing a marker for studying brain aging and cognitive specialization [84].

The richness of the information captured by the edge-centric paradigm has catalyzed a surge of successful applications in clinical and cognitive neuroscience, consistently demonstrating its power in prediction and biomarker discovery. A primary strength of the framework lies in its superior ability for subject identification and capturing individual differences. Studies have shown that eFC provides a more robust and unique fingerprint of an individual's brain compared to node-centric methods [85] and crucial for developing personalized network maps [86]. This sensitivity to individual variability has translated into powerful models for predicting behavioral and cognitive traits, such as economic risk-taking [87], loss aversion in internet gaming disorder [88], and transdiagnostic factors of depression and anxiety [89].

The edge-centric paradigm has emerged as a highly effective tool for biomarker discovery across a wide spectrum of neurological and psychiatric conditions. It has been instrumental in characterizing network alterations in neurodevelopmental disorders like autism spectrum disorder [90] and in infants with prenatal opioid exposure [74]. Similarly, it has provided novel insights into the pathophysiology of neurological conditions such as depression and anxiety comorbidity [89], [91], stroke [92], multiple sclerosis [93], migraine [94], and early-stage mild cognitive impairment (eMCI) [95]. Furthermore, the framework has been applied to uncover the fundamental principles of brain organization and development, for instance by studying the temporal organization of specific circuits like corticostriatal pathways [96], and the longitudinal development of overlapping functional modules throughout childhood and adolescence [97].

The foundational concepts of the edge-centric paradigm have also inspired several methodological extensions that broaden its scope and applicability. One significant generalization moves beyond temporal co-fluctuation by defining an edge feature vector, which can encode the strength of a connection across multiple tasks, conditions, or even imaging modalities. The covariance of these feature vectors then provides a measure of higher-order relationships in multi-relational datasets [98]. In a similar spirit, Yang et al. (2023) extended the focus from temporal dynamics to population dynamics by defining the key feature of an edge as its inter-individual variability. Their work revealed a 'connectional hierarchy' in the brain, demonstrating the flexibility of the edge-centric philosophy to capture different facets of network organization [99].

*2.1.1.3. Metric-based Triplet Correlation*

A conceptually related, yet methodologically distinct, approach to capturing second-order relationships is through metric-based triplet correlation. Instead of comparing the global connectivity profiles central to topological HOFC, this method focuses on the local geometry of interactions within triplets of brain regions. It defines a higher-order correlation between two regions by quantifying the similarity of their distance patterns to a shared set of immediate neighbors, using the squared Euclidean distance between their time series as the base metric. Similar to the CoC framework, this process results in a new second-order connectivity matrix that captures beyond-pairwise information. This local, metric-based approach has been shown to be particularly effective in identifying subtle network alterations, and its fusion with first-order connectivity networks has demonstrated enhanced performance in the classification of schizophrenia [100].

## 2.1.2. Multivariate Statistical Dependencies

While the CoC framework infers higher-order relationships by analyzing the similarity of patterns of pairwise connectivity, a more direct paradigm involves quantifying the simultaneous statistical dependence among a group of three or more neural time series. This broad family of methods, rooted in the principles of multivariate statistics, seeks to capture irreducible group interactions that are not fully described by the sum of their constituent pairwise relationships. This section is organized into two major sub-paradigms that tackle this challenge from different theoretical perspectives. First, we will review a diverse set of methods derived from classical statistics and signal processing, including those based on modeling probability distributions [55], capturing higher-order statistical moments [101], and measuring multivariate phase synchronization [102]. Second, we will delve into the comprehensive and model-free framework of information-theoretic measures, which provides a principled language for decomposing complex dependencies into fundamental concepts such as redundancy and synergy [6], [103], [104].

*2.1.2.1. Matrix Variate Normal Distribution*

An important approach within this domain is the MVND framework, which holistically models the entire functional connectivity (FC) matrix as a single random statistical object. The method begins by generating a population of dynamic FC matrices, typically using a sliding-window [55], [105], [106] approach on time series data. This collection of dynamic FC matrices is then conceptualized as samples drawn from an underlying MVND. This elegant formulation allows for the simultaneous estimation of both low- and higher-order connectivity from the distribution's parameters [55]. Conceptually, this approach is closely related to the dynamic HOFC (dHOFC) method [11], as both seek to capture the co-fluctuation of pairwise connections over time [11], [55]. However, instead of using sequential correlations, the MVND framework leverages a principled statistical distribution where the mean matrix (M) represents the stable, average pairwise connections (low-order FC), and the covariance matrix (Ω) directly quantifies the higher-order relationships, or the correlations between the edges themselves [55].

The principled statistical foundation of MVND has spurred a cascade of advancements aimed at refining its robustness and theoretical rigor. Recognizing that the initial higher-order networks were often dense and noisy, Zhou et al. incorporated regularization terms into the estimation process. By adding penalties that enforce sparsity (promoting fewer, more significant connections) and modularity (favoring biologically plausible community structures), they produced more interpretable and reliable higher-order networks [107]. Further strengthening its theoretical basis, Jiang et al. proposed a Bayesian reformulation (BHM), which treats the low- and higher-order networks as latent parameters of a unified generative model, enabling more rigorous estimation [108]. To overcome the computational challenges of applying MVND to large-scale brain networks, a hierarchical strategy was developed, which partitions the brain into smaller subnetworks before integrating the results [109]. The framework's versatility is also notable, having been successfully adapted to other modalities like EEG for identifying higher-order biomarkers in depression [106] and cognitive-motor tasks [105].

*2.1.2.2. Multivariate Cumulants*
While the MVND framework primarily leverages second-order moments (covariance), other methods directly probe higher-order statistical moments to capture non-Gaussian and nonlinear dependencies. A prime example is the use of multivariate cumulants. The key advantage of cumulants, such as third-order coskewness and fourth-order cokurtosis, is their ability to isolate genuine higher-order interactions by design. They inherently disregard statistical dependencies that can be explained by lower-order (i.e., pairwise) correlations, thus providing a measure of irreducible, non-redundant group interactions. Hindriks et al. introduced a robust framework for estimating these multivariate cumulants from fMRI data, complemented by a block bootstrapping procedure for rigorous statistical inference, demonstrating their utility in characterizing the resting-state brain and classifying clinical populations [101].

*2.1.2.3. Multivariate Phase Synchronization*
For neural data with non-stationary signals, such as EEG and MEG, analyzing dependencies in the phase domain offers unique insights into neural coordination. Moving beyond pairwise measures, multivariate phase synchronization methods aim to quantify the simultaneous phase-locking among three or more signals. A particularly innovative approach in this area is the Symbolic Phase Difference and Permutation Entropy (SPDPE) framework, a non-linear and robust method for measuring global phase synchronization [102], [110]. The core of SPDPE lies in its unique definition of phase interaction. For each signal in a group, it calculates the phase difference relative to a reference phase derived from the circular mean of all other signals in that group. This captures how each oscillator aligns with the collective rhythm of its ensemble. These continuous phase differences are then converted into a sequence of discrete symbols, and the complexity of this symbolic sequence is quantified using permutation entropy. A low entropy (SPDPE value near 1) indicates a highly ordered and predictable phase relationship, signifying strong global synchronization, while high entropy (SPDPE value near 0) suggests desynchronization. This approach has proven highly effective in capturing the global synchronization dynamics during epileptic seizures [102], [110].

*2.1.2.4. Information-Theoretic Measures*
Information theory, pioneered by Shannon, offers a powerful, inherently non-linear, and model-free mathematical framework for quantifying statistical dependencies. While its foundational concept, Mutual Information (MI), is defined for pairs of variables, its generalization to multivariate systems has given rise to an ecosystem of measures for probing higher-order interactions in the brain [111]. Figure 4 provides a conceptual illustration of several key multivariate information measures using Venn diagrams. This evolution has been driven by the need to understand two fundamental, yet distinct, aspects of multi-way statistical relationships: redundancy, the extent to which information is duplicated or shared across a system, and synergy, the creation of novel information that emerges only from the interaction of a group as a whole and is irreducible to its constituent parts [112], [113]. The quest to quantify and disentangle these two phenomena has led to a historical and conceptual progression of increasingly sophisticated measures [6], [103], [104], [112], [114], [115], [116].

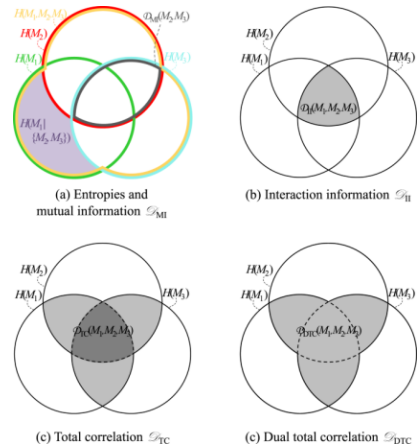

*Figure 4. Information-theoretic measures of higher-order interactions illustrated with Venn diagrams for a system of three variables ($M_1$, $M_2$, $M_3$) [117]. (a) Foundational Concepts: The area of each circle represents the marginal entropy of a single variable (e.g., $H(M_1)$). The union of circles represents the joint entropy of the system ($H(M_1, M_2, M_3)$), while overlaps represent mutual information between subsets (e.g., $D\_MI(M_2, M_3)$). Conditional entropy (e.g., $H(M_1|\{M_2, M_3\})$) is represented by the unique area of a circle. (b) Interaction Information (II): This measure quantifies the net balance between redundancy and synergy in a triplet. It corresponds to the central intersection area, representing the information shared by all three variables simultaneously. A positive value indicates redundancy, while a negative value indicates synergy. (c) Total Correlation (TC): Also known as multi-information, TC measures the total amount of redundant information in the system. It quantifies the information shared by at least two variables and is represented by the sum of all pairwise and tripletwise intersection areas. (d) Dual Total Correlation (DTC): This measure quantifies the amount of information that is unique to the system as a whole and cannot be found in any single variable. It is represented by the central intersection area plus the areas unique to each pairwise intersection. These measures form the basis for other decompositions. For instance, the O-information (Ω), a key metric for distinguishing redundancy- from synergy-dominated systems, is defined as the difference between Total Correlation and Dual Total Correlation (Ω = TC - DTC).*

The earliest information-theoretic approaches to quantifying group dependency were built upon the foundational concept of Shannon entropy. For a group of variables, the multidimensional entropy measures the total uncertainty or unpredictability of the system's collective state. A lower joint entropy implies stronger statistical interdependencies, as the state of the system as a whole is more predictable. This principle was powerfully leveraged by the CASH algorithm, which partitions brain networks into functional ensembles by finding groups of regions that minimize their joint entropy, thereby identifying maximally co-regulated systems [118]. Building directly on multidimensional entropy, Total Correlation (TC), or multi-information, was formulated to isolate the portion of this dependency attributable to shared information or redundancy [119]. Defined as $TC(X) = \Sigma H(X_i) - H(X)$, it quantifies the difference between the information needed to describe each variable independently (the sum of individual entropies) and the information needed to describe the entire system jointly. Conceptually, this difference represents the informational overlap or the amount of information that is duplicated across the variables, providing a direct measure of the group's overall redundancy.

In higher-order network neuroscience, TC is typically employed not as a single metric for the entire brain, but as a tool to systematically probe interactions within smaller subsets of brain regions, most commonly triplets [120], [121], [122], [123]. By computing the TC for all possible triplets of neural time series, researchers can construct a high-dimensional tensor of interaction strengths, revealing which specific three-way interactions exhibit the highest redundancy [120], [123]. This approach provides a rich, system-wide map of higher-order dependencies that has been used to compare network organization across different cognitive states [122] and between different modeling approaches [121]. Moreover, TC has been leveraged as a multivariate distance metric for hierarchical clustering algorithms, enabling the data-driven discovery of functional communities that are more consistent with underlying anatomical structure than those derived from pairwise measures [124]. Furthermore, the values derived from TC have served as a principled, data-driven basis for constructing explicit higher-order structures. In this hybrid paradigm, the triplets with the highest TC values are selected to form the fundamental building blocks, either as hyper edges in a hypergraph [103], [125] or as 2-simplices in a simplicial complex [126], thus bridging the gap between implicit information-theoretic measures and explicit structural models.

While TC quantifies the redundancy within a system, the complementary concept of Dual Total Correlation (DTC) was developed to measure its integration [127]. DTC quantifies the amount of information encoded in the system as a whole that is not present in any of its constituent parts. Theoretically, it measures the difference between the sum of the information of all individual variables and the information of the entire system when partitioned into n-1 variables. A high DTC value signifies that the system is highly integrated, meaning that its collective state contains significant information that cannot be understood by observing its components in isolation [128]. This makes it a powerful metric for identifying functionally integrated neural ensembles. For instance, Herzog et al. (2022) applied this concept to resting-state fMRI data, computing the DTC for all possible triplets of brain regions. By identifying the triplets with the highest integration values, they were able to construct higher-order brain networks that served as effective biomarkers for distinguishing patients with frontotemporal dementia and Alzheimer's disease from healthy controls [6].

The realization that TC and DTC capture distinct, often opposing, aspects of multivariate dependency, redundancy versus integration, led to the development of methods that capture both aspects. An early and influential attempt in this direction was Interaction Information (II), which quantifies the net balance for a triplet of variables [129]. Defined as II(X,Y,Z) = MI(X,Y) - MI(X,Y|Z), it measures how the presence of a third variable Z modulates the relationship between X and Y. Positive II is interpreted as redundancy, while negative II signifies synergy. This measure was successfully used to study the lifespan dynamics of these interactions in the resting brain [130]. However, the limitations of II, particularly its restriction to triplets and the ambiguity of its interpretation for larger groups, motivated the development of a more general and scalable solution: the O-information ($\Omega$) framework [128].

By defining $\Omega$ = TC - DTC, O-information provides a single, scalable metric that quantifies the net balance between redundancy (indicated by $\Omega > 0$) and synergy (indicated by $\Omega < 0$) for a group of variables of any size. This breakthrough resolved many of the earlier measures and provided a direct tool to probe the nature of higher-order interactions. In higher-order neuroscience, this is typically achieved by computing the O-information for a vast number of small subsets (e.g., triplets or quadruplets) of brain regions to create a distribution of interaction patterns in fMRI [104], [131], EEG [132], [133] or MEG [132]. This approach has yielded a wide array of insights, from characterizing age-related shifts towards redundancy in brain dynamics [104] and identifying aberrant synergistic/redundant patterns in schizophrenia [115], to uncovering synergistic shadow structures in the connectome that are invisible to pairwise methods [131], [134]. To provide a more complete picture, the non-negative S-information was introduced as a complementary measure that quantifies the net amount of synergistic information, isolating the computational complexity of a system from its co-existing redundant components [115]. The framework continues to evolve with extensions such as the Spectral O-information Rate for frequency-domain data [135] and more robust estimators for non-Gaussian data [132].

A bold achievement of this pursuit is the Partial Information Decomposition (PID) framework, which aims to provide a complete and non-negative decomposition of multivariate information [136], [137]. While theoretically challenging, applied PID frameworks have begun to unravel the distinct roles of higher-order interactions. These methods decompose the information that multiple sources provide about a target into unique, redundant, and synergistic components. For example, by applying PID to fMRI data, Luppi et al. (2022) demonstrated that synergistic and redundant interactions follow distinct neuroanatomical hierarchies, with synergy dominating in association cortices and redundancy in sensorimotor areas [112]. Other variants, such as Partial Entropy Decomposition (PED) [138], [139] and Integrated Information Decomposition (IID) [140], have further illuminated the time-resolved dynamics of these information modes and their alteration in psychiatric disorders. These advanced decomposition frameworks represent the current frontier in providing a complete and interpretable picture of how the brain integrates and computes information through complex, multi-way interactions.

### 2.1.3. Graph-Based Paradigms

A final category of implicit methods moves beyond statistical distributions [55], [107], information theory [6], [125] or high-order correlation [11], [12] or statistics [141] to infer higher-order interactions from dynamic processes unfolding on a graph structure. These innovative approaches, inspired by dynamic graph theory [142] and statistical physics [143], model phenomena such as network traversal or temporal asymmetry to reveal complex, global relationships that are not captured by local statistical measures.

A method infers global, higher-order relationships by modeling processes unfolding on an underlying graph structure. The Signed Random Walk framework, for instance, models the traversal of a random walker on a signed functional network, incorporating principles from structural balance theory to handle negative connections. The final relationship between two nodes is not their direct edge, but an implicit measure derived from the probabilities of all possible paths connecting them, thus embedding global network structure into a new higher-order connectivity matrix [142].

Another conceptually distinct approach, inspired by statistical physics, focuses on non-equilibrium dynamics by first transforming neural time series into complex networks using the Visibility Graph (VG) algorithm [144]. The VG provides a principled way to map a time series to a graph, where each time point is a node and an edge is drawn between two time points if their corresponding values in the time series plot are mutually visible (i.e., the straight line connecting them is not obstructed by any intermediate value) (Fig. [Number]). While this powerful tool is also used in the explicit modeling paradigm to construct Simplicial Complexes from the resulting graph's cliques [145], its application here follows a different philosophy. Nartallo-Kaluarachchi et al. extended this concept to directed, multiplex VGs to quantify the degree of temporal irreversibility. Their higher-order measure captures the collective contribution of a group of brain regions (k-tuple) to the breaking of time-reversal symmetry by comparing the in- and out-degree distributions on the graph using Jensen-Shannon Divergence. This reveals a novel form of implicit interaction related to the directional flow of information and emergent computation, demonstrating how a single foundational tool like the VG can be leveraged for both implicit quantification and, as will be discussed later, explicit structural construction [143].

### 2.2. Explicit Methods

Having surveyed the methods that quantify group dependencies implicitly, we now turn to the explicit paradigm. These frameworks are fundamentally concerned with constructing and analyzing new mathematical objects, namely hypergraphs and Simplicial Complexes, that directly represent group interactions. This conceptual shift moves the analytical challenge from statistical estimation to the construction and topological characterization of these objects.

A critical consequence of this shift to explicit structures is the emergence of two distinct but intertwined research thrusts: construction and analysis. Unlike the implicit methods, where the analysis is often inherent in the extraction metric itself, the explicit paradigm necessitates a subsequent step to interpret the complex objects that have been built. Consequently, the literature has evolved along two parallel paths: developing novel methods to construct biologically plausible hypergraphs and Simplicial Complexes, and designing powerful analytical tools to extract meaningful features from these structures. Intriguingly, this creates a symbiotic relationship between the implicit and explicit paradigms, where statistical metrics from the former can serve as powerful tools for the latter. Intriguingly, this creates a symbiotic relationship between the paradigms, where statistical metrics from implicit approaches can serve as powerful, data-driven tools for constructing or weighting these explicit structures [126]. As we will also see, particularly in modern deep learning architectures, the boundary between construction and analysis can become further blurred, with both processes often being optimized jointly in an end-to-end fashion [146], [147], [148].

At the forefront of this explicit modeling paradigm are two complementary, yet philosophically different, formalisms. The first is the hypergraph, as shown in Figure 5b, which provides a highly flexible, set-based language for describing group interactions [149], [150], [151]. Conceptually, a hypergraph can be likened to a list of scientific co-authorships, where each paper (a hyper edge) defines a specific collaborating group of researchers (nodes), without imposing constraints on the relationships within subgroups. The second formalism is the simplicial complex (Figure 5a), a more structured and hierarchical framework originating from algebraic topology [152], [153], [154]. This approach focuses on uncovering the intrinsic shape and topology of neural data. Drawing an analogy, if hypergraphs identify functional committees, Simplicial Complexes map the entire organizational landscape, revealing not only clusters but also the voids, loops, and cavities that shape information flow [153], [154]. This section will systematically explore the methodologies developed for building and analyzing these two powerful structures, elucidating their unique contributions to understanding the higher-order organization of the brain.

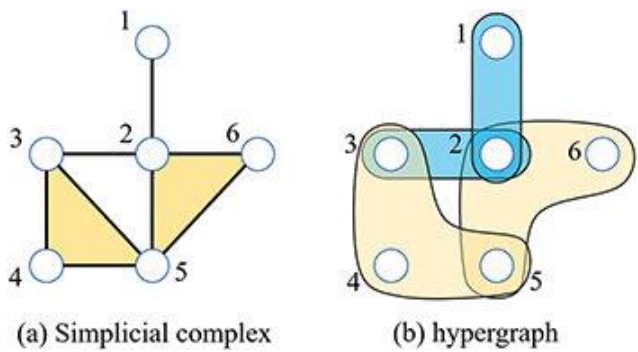

Figure 5. A comparison of Explicit Models for Higher-Order Interactions: Simplicial Complexes vs. Hypergraphs [155]. Both structures generalize traditional graphs to represent interactions among more than two nodes, but they do so with different underlying philosophies. (a) Simplicial Complex: A simplicial complex is a hierarchical structure built from geometric primitives called simplices (nodes, edges, filled triangles, etc.). A key constraint is the "face rule": if a simplex (e.g., the filled triangle {3,4,5}) exists, all its lower-dimensional faces (the edges {3,4}, {4,5}, {3,5} and the nodes {3},{4},{5}) must also exist. This structure is ideal for topological data analysis and studying the "shape" of data. (b) Hypergraph: A hypergraph is a set-based structure where a hyperedge can connect any number of nodes, without imposing constraints on the relationships between subsets of nodes. For example, the yellow hyperedge connects nodes {2,3,4,5,6} as a single group, even though not all nodes within this group are connected to each other. This flexible framework is well-suited for modeling group membership and community structures.

### 2.2.1. Hypergraphs: Modeling Set-Based Interactions

The hypergraph provides a natural and powerful generalization of the traditional graph formalism, designed to explicitly capture polyadic, or set-based, interactions. While a standard graph is restricted to representing pairwise relationships through its edges, a hypergraph relaxes this constraint by introducing the concept of a hyper edge: a connection that can link any number of nodes simultaneously [156]. Formally, a hypergraph H is defined by a pair (V, E), where V is a set of nodes (e.g., brain regions) and E is a set of hyper edges, with each hyper edge being a non-empty subset of V. This structure is typically represented by an incidence matrix, H, where the entry H(i, j) is 1 if node i is a member of hyper edge j, and 0 otherwise. The inherent flexibility of this representation allows for the modeling of diverse group activities, from small, tightly-knit functional motifs to large-scale brain networks, without imposing the rigid hierarchical assumptions required by other topological structures.

The primary challenge in applying hypergraphs to neuroscience, however, lies in a fundamental question: how should one define and identify neurobiologically meaningful hyper edges from complex, high-dimensional neural data? The literature has converged on two principal philosophies for addressing this construction problem, which can be broadly categorized as model-driven and data-driven paradigms. Model-driven approaches frame hyper edge identification as a statistical inference or signal reconstruction problem. They typically assume an underlying generative model—most commonly, a sparse linear regression—and seek the most parsimonious set of brain regions that can explain the activity of a target region [149], [157], [158], [159]. In contrast, data-driven approaches are non-parametric and operate directly on the observed data to identify groups of nodes that exhibit high similarity, whether based on their temporal activity [160], connectivity profiles [161], or pre-defined anatomical knowledge [162], [163]. These two paradigms, while distinct in their methodological underpinnings, offer complementary perspectives on uncovering the rich tapestry of multi-way interactions that constitute brain function.

#### 2.2.1.1. Model-Driven Hypergraph Construction

The model-driven paradigm for hypergraph construction was pioneered in neuroscience by Jie et al. (2016), who introduced a seminal framework based on sparse linear regression. This approach elegantly reframes hyper edge identification as a feature selection problem: for each brain region, it seeks the smallest subset of other regions whose activity can best reconstruct its time series. The core of this method lies in solving a Lasso-regularized optimization problem for each target region m:

$$\min \|x\_m - A\_m * \alpha\_m\|^2 + \lambda * \|\alpha\_m\|_1.$$

The l1-norm penalty $\|\alpha\_m\|_1$ enforces sparsity on the coefficient vector α_m, effectively selecting a parsimonious set of predictive regions. A hyper edge is then defined as the union of the target region m and all predictor regions with non-zero coefficients. By iterating this process over all brain regions, a comprehensive hypergraph representing the brain's hyper-connectivity is constructed [149]. This foundational framework quickly demonstrated its utility in clinical applications. Its applications span multiple imaging modalities and conditions, including the characterization of Moyamoya disease based on cognitive profiles [164], AD classification [165] identifying biomarkers for MCI, Major Depressive Disorder (MDD) [166] and ADHD from fMRI data [149], and uncovering higher-order patterns in EEG data for diagnosing major depressive disorder [167]. This versatility has also been leveraged for tasks like discovering network phenotypes linked to genetic risk [168].

A critical limitation of the standard Lasso formulation, however, is its instability in the presence of highly correlated predictors; a ubiquitous feature of neuroimaging data. This grouping effect causes Lasso to arbitrarily select one region from a functionally cohesive group while discarding others. To address this, subsequent work introduced methods like Elastic Net and Group Lasso (also known as the l1,2-norm sparse algorithm) to encourage the selection of correlated brain regions as a group [157], [169], [170]. This line of continued in the adoption of Sparse Group Lasso (sgLasso), a more sophisticated model that performs a bi-level selection: first at the group level and subsequently at the individual region level within selected groups. This refinement allows for a more neurobiologically plausible selection of functional ensembles, leading to more robust hypergraph representations for tasks such as MDD classification [158].

The initial static and undirected nature of these models spurred further extensions to capture the brain's dynamic and directional properties. The integration of sparse regression with sliding-window analysis enabled the construction of dynamic hyper networks, capable of tracking temporal fluctuations in higher-order functional interactions [159]. In another interesting approach, Lasso-based hypergraphs were employed as the highest layer in a complex spatio-temporal multilayer network by combining both white and gray matter data to investigate the dynamic organization of the brain's rich-club structure [171]. A significant conceptual advance was the introduction of directionality. By incorporating a temporal lag effect into the sparse regression model, a framework for inferring Effective Hyper-Connectivity (EHC) were developed. This innovation shifted the paradigm from modeling symmetric associations to inferring directed, causal-like information flow within higher-order circuits, which was then analyzed using a Directed Hypergraph Convolutional Network (DHGCN) [172].

Beyond refining the core construction process, recent work has focused on more sophisticated methods for analyzing and leveraging the constructed hypergraphs, often blurring the line between construction and analysis. An approach is the development of novel hyper edge weighting schemes. In this approach, after constructing a hypergraph using Lasso, each hyper edge is weighted by its algebraic connectivity, a holistic measure of its internal coherence that moves beyond simple pairwise metrics [173]. Another innovative direction involves integrating the hypergraph structure directly into machine learning pipelines. This study developed a multi-attribute hypergraph kernel that computes the similarity between entire brain hypergraphs by considering both their higher-order topology and the functional attributes of their nodes. Perhaps the most advanced integration is seen in end-to-end deep learning frameworks [174]. This method proposed a dynamic weighted hypergraph convolutional network (dwHGCN) where the hyper edge weights are not predetermined but are adaptively learned during model training, allowing the framework to dynamically emphasize the most task-relevant higher-order interactions [146]. These developments illustrate a clear trajectory in the field: from a fixed, two-step process of construct then analyze towards unified frameworks where the hypergraph structure itself is a dynamic, learnable, and integral component of the analytical model.

### *2.2.1.2. Data-Driven Hypergraph Construction*

In contrast to model-driven methods that rely on an underlying statistical model, data-driven approaches offer a non-parametric and flexible alternative for hypergraph construction [147], [161]. This paradigm operates directly on the intrinsic properties of the neural data, defining hyper edges based on observed patterns of similarity [161], proximity [160], or shared information [148], without assuming a specific generative process like linear regression. This diverse family of methods can be further organized by the principle guiding their construction, ranging from local neighborhood definitions and clustering to the integration of prior knowledge and sophisticated end-to-end learning schemes.

A foundational strategy within the data-driven paradigm is based on neighborhood and clustering principles, where hyper edges are formed by grouping brain regions that exhibit similar characteristics. In its most direct form, this involves defining a neighborhood for each brain region based on functional connectivity. For example, the Good Friends approach, first constructs a standard pairwise functional connectivity network and then defines a hyper edge for each region as the set containing the region itself and all other regions to which it is connected above a certain correlation threshold [161]. Similarly, Du et al. (2025) construct their initial weighted hypergraph by defining hyper edges based on the first-order neighborhood of each node in the underlying brain network [151]. A closely related concept employs clustering algorithms like k-nearest neighbors (KNN) to define hyper edges based on local similarity in the feature space of brain regions [160], [175]. This principle can also be applied at the subject level, where a hypergraph is constructed to group individuals with similar subnetwork connectivity patterns, thereby identifying disease-related biomarkers directly from population data [150].

Expanding on this, prior-informed and hybrid methods enrich the construction process by incorporating external knowledge, leading to more neurobiologically constrained hypergraphs. Instead of relying solely on functional data, these approaches leverage anatomical or established functional atlases as a scaffold for defining hyper edges. For instance, Zuo et al. utilized a prior-guided approach within a deep learning framework to construct hypergraphs that respect known brain organization [162], [176]. Another particularly compelling example is the explicit modeling of structure-function coupling, where structural connectivity (SC) data from diffusion imaging is used to inform the construction of a functional hypergraph. Ma et al. (2021) developed a hypergraph Laplacian diffusion model where hyper edges are defined by the underlying structural pathways, providing a powerful tool to predict functional correlations from the brain's anatomical wiring [177]. Similarly, Han and Li (2025) integrated prior knowledge of established resting-state networks as one view in their multi-view hypergraph learning framework, combining it with purely data-driven views to create a richer representation [163].

A distinct and principled data-driven strategy for hyperedge construction leverages the power of information-theoretic measures to identify statistically significant group interactions. Instead of relying on similarity or neighborhood definitions, this approach quantifies the strength of higher-order dependencies using multivariate information metrics. For instance, both Rawson (2022) and Santos et al. (2023) utilized TC, a measure of a group's total redundancy, to define and weight hyper edges. In these frameworks, the TC is computed for all possible triplets of brain regions, and the most informative groups (i.e., those with the highest TC values) are selected to form the hyper edges of the brain's entropic hyper-connectome. This method provides a rigorous, non-parametric way to identify multi-node interactions, which can then be analyzed to discover higher-order functional hubs or used as features for clinical classification [103], [125].

The more recent and rapidly evolving frontier in data-driven construction involves end-to-end learning-based methods, where the hypergraph structure is not a fixed, predefined entity but a set of learnable parameters optimized jointly with a downstream task, such as disease classification. These deep learning frameworks treat the hypergraph itself as part of the model architecture to be discovered. For example, Qiu et al. (2023) and Hu et al. (2025) proposed models that learn binary masks over brain regions to identify the most informative and minimally redundant hyper edges for a given cognitive or clinical prediction task, often guided by principles from information theory. This data-driven optimization allows the model to discover novel, task-relevant higher-order circuits that might be missed by methods based on fixed construction rules [147], [148]. Further advancing this integration, some frameworks combine the strengths of both paradigms; for instance, the dynamic weighted hypergraph convolutional network (dwHGCN) proposed by Wang et al. (2023) uses a sparse regression model for initial hyper edge proposal but then adaptively learns the weights of these hyper edges in an end-to-end fashion. This trend towards learnable hypergraphs represents a new paradigm, transforming hypergraph construction from a preprocessing step into a dynamic and integral part of the data analysis pipeline itself [146].

*2.2.1.3. Analysis of Hypergraph Structures*

Once a hypergraph representation of the brain network is constructed, the subsequent challenge lies in extracting neurobiologically meaningful information from its complex, higher-order topology. The analytical tools developed for this purpose range from direct generalizations of classical graph metrics [149] to sophisticated, learning-based frameworks that leverage the full richness of the hypergraph structure [147], [163]. These methods provide a new vocabulary for characterizing brain organization, moving beyond pairwise connections to quantify group-level phenomena.

A primary approach to hypergraph analysis involves the generalization of classical graph-theoretic metrics. These tools provide an intuitive first step for quantifying the properties of the constructed higher-order network. Foundational metrics include the node degree, defined as the number of hyper edges a node participates in, and the hyper edge size (or cardinality), which simply counts the number of nodes within a hyper edge [157]. More complex measures, such as the hypergraph clustering coefficient, have also been proposed to characterize local network topology. Several variants of this metric exist, designed to capture different aspects of local connectivity, for instance, by quantifying the extent of overlap between the hyper edges incident to a given node [149], [165].

Another powerful analytical technique involves kernel methods, which offer a way to compare entire hypergraph structures without explicit feature vectorization. By designing a kernel function, one can compute a similarity score between the brain hypergraphs of two different subjects, enabling their direct use in classifiers like Support Vector Machines (SVMs). One notable example is the multi-attribute hypergraph kernel, which considers both topology and nodal attributes [174]. A particularly innovative approach in this domain is the Hyper-Ordinal Pattern (HOP) framework. Instead of relying on static topological features, this method defines a unique signature for each node by tracing a path through its incident hyper edges based on the descending order of their weights. This process generates a hyper-ordinal pattern—a topological-sequential signature that captures not just group membership, but also the relative importance of the different functional communities connected to a given brain region. A specific kernel is then designed to measure the similarity between these HOPs, providing a rich, dynamics-aware method for network comparison [151].

Alternative analytical approaches involve transforming the hypergraph into other structures to reveal different topological properties. For example, the high-order line graph method converts the original hypergraph (where nodes are brain regions and edges are groups of regions) into a new graph where nodes represent the original hyperedges [166].

One of the most significant recent advances in hypergraph analysis, however, have emerged from the integration with deep learning, leading to powerful end-to-end frameworks. At the heart of this trend are Hypergraph Neural Networks (HGNNs), which generalize the message-passing mechanism of standard Graph Neural Networks to higher-order structures. In an HGNN, information is aggregated not just from pairwise neighbors, but from all nodes within a shared hyper edge, allowing the model to learn complex, multi-way feature representations directly from the hypergraph topology [163], [169]. This integration has led to a paradigm shift where the processes of construction and analysis become deeply intertwined and often co-optimized. For example, in the dynamic weighted HGCN (dwHGCN) framework, an initial hypergraph is constructed via sparse regression, but the crucial hyper edge weights, which determine the strength of higher-order interactions, are adaptively learned during the training of the neural network to maximize task performance [146]. Similarly, other frameworks treat the very structure of the hypergraph, i.e., the membership of nodes in hyper edges, as a set of learnable parameters, using techniques like learnable masks to discover the most informative and minimally redundant hyper edges for a given predictive task [147], [148]. This move towards learnable, task-optimized hypergraphs represents the current frontier of the field, transforming the hypergraph from a static data representation into a dynamic and integral component of the computational model itself.

Beyond message-passing frameworks, novel analytical tools inspired by signal processing are also emerging. For instance, Hypergraph Signal Processing, using techniques like the Hypergraph Fourier Transform based on tensor decomposition, allows for a spectral analysis of the constructed high-order network, revealing frequency-based differences between clinical populations [170].

## 2.2.2. Simplicial Complexes and Topological Data Analysis: Discovering the Shape of Interactions

While hypergraphs provide a flexible, set-based framework for modeling group interactions, a parallel and mathematically rich paradigm has emerged from the field of algebraic topology: Topological Data Analysis (TDA). The central premise of TDA is to study the intrinsic shape of data by representing it as a simplicial complex [152]. Unlike a hypergraph, which is fundamentally a collection of sets, a simplicial complex is a more constrained, hierarchical structure built from geometric primitives called simplices (nodes, edges, filled triangles, solid tetrahedra, and their higher-dimensional counterparts). This hierarchical rule, mandating that if a simplex exists, all its lower-dimensional faces must also exist, imposes a strong topological consistency. This distinction is crucial: In contrast to the set-based representation of hypergraphs, Simplicial Complexes model the entire organizational 'landscape' of the data. This approach allows for the characterization of not just clusters of activity, but also the loops, voids, and cavities that shape the relationships between them [154].

The fundamental building blocks of a simplicial complex are simplices, which generalize the notions of points, lines, and triangles to higher dimensions [152], [178]. A 0-simplex is a node (a vertex), a 1-simplex is an edge connecting two nodes, a 2-simplex is a filled triangle defined by three fully interconnected nodes, and a 3-simplex is a solid tetrahedron formed by four fully interconnected nodes. The crucial distinction lies in their interpretation: a 2-simplex, for example, represents not merely three pairwise relationships, but a single, irreducible tripartite interaction [152]. A collection of such simplices forms a simplicial complex only if it adheres to a strict hierarchical consistency rule: if a simplex is part of the complex, then all its lower-dimensional faces (e.g., the edges and vertices of a triangle) must also be included. This face rule ensures that the resulting structure is topologically coherent, providing a valid scaffold upon which the tools of algebraic topology can be applied to study its shape [178].

### *2.2.2.1. Conceptual Foundations and Construction Methods*

The most prevalent method for constructing a simplicial complex from neural data begins with a pairwise connectivity graph, a process known as clique complex construction [153], [179]. This approach first computes a weighted connectivity matrix (e.g., via Pearson correlation) across all brain regions. To capture multi-scale topological features and circumvent the arbitrary choice of a single connectivity threshold, a filtration is typically performed [152], [154]. This involves creating a nested sequence of graphs by progressively adding edges in order of their weights, from strongest to weakest. At each step of this filtration, the resulting graph is promoted to a clique complex: every k-clique (a set of k nodes that are all mutually connected) in the graph is treated as a (k-1)-simplex. For instance, a fully connected triplet of nodes (a 3-clique) is not merely represented as three edges, but as a filled-in 2-simplex (a triangle). This process generates a sequence of nested Simplicial Complexes, providing a rich, multi-scale representation of the data's topological evolution, which serves as the input for persistent homology analysis [179], [180].

An alternative and innovative approach bypasses the need for an initial connectivity matrix by constructing a simplicial complex directly from the dynamics of individual time series. A notable example of this is the Visibility Algorithm, which transforms a time series into a graph by connecting time points that are mutually visible along the signal's trajectory [144]. In this framework, each time point is a node, and an edge exists between two time points if the straight line connecting them is not obstructed by any intermediate data point. This graph, which captures the geometric and fractal properties of the time series, is then converted into a clique complex. This technique, applied to EEG [181] and fMRI [145] signals, allows for the characterization of the local, dynamic complexity of individual brain regions, which can then be compared across different cognitive states or clinical populations [145], [181].

### 2.2.2.2. Analysis of Simplicial Structures

Once a simplicial complex is constructed, it provides a rich topological scaffold for uncovering higher-order organizational principles that are invisible to standard graph metrics. The analysis of these structures moves beyond simply counting components to characterizing their intrinsic shape, most notably through the identification of topological features such as cycles, voids, and cavities [152], [153]. A cycle, or a 1-dimensional hole, can represent a recurrent information processing loop that lacks a central hub, while higher-dimensional cavities may signify functional segregation or boundaries between distinct neural assemblies. In the following will detail the three primary analytical paradigms, combinatorial [153], algebraic [153], [179], and signal processing [126]-based, that have been developed to quantify these and other complex topological properties.

The most direct analytical approach, rooted in combinatorial topology, involves characterizing the simplicial complex by enumerating its constituent building blocks. This method provides a computationally efficient, global summary of the network's higher-order complexity. The primary tool for this is the f-vector, $f = [f_0, f_1, f_2, ... ]$, which counts the total number of k-simplices in the complex and has been used to compare brain dynamics across populations and conditions [145], [181]. A related approach, Q-analysis, focuses specifically on the structure of maximal cliques and their intersections to reveal the hierarchical organization of the network [182]. This principle of using maximal cliques as fundamental higher-order units has been applied to EEG data to identify hyper-networks in boys with ADHD. In this work, after detecting maximal cliques using the Bron-Kerbosch algorithm, topological features derived from these structures, such as hyper-network clustering coefficients, were analyzed to reveal group differences [183]. A more sophisticated global invariant derived from this is the Euler characteristic ($\chi$), calculated as the alternating sum of the f-vector components ($\chi = f_0 - f_1 + f_2 - ...$). By tracking the Euler characteristic across a network filtration, Ling et al. (2025) demonstrated that its evolution reveals distinct topological phase transitions in brain functional networks, identifying critical points of global reorganization that correlate with changes in classical graph metrics and are altered in Parkinson's disease [184]. This combinatorial approach has also been used to analyze Simplicial Complexes constructed from multivariate phase synchronization measures, demonstrating how implicit metrics can inform the analysis of explicit structures [110].

While combinatorial methods count the components of a simplicial complex, algebraic topology provides the machinery to understand how these components assemble to form a global shape [152], [154]. This paradigm focuses on identifying and quantifying topological invariants like connected components, cycles (1-dimensional holes), and voids (higher-dimensional cavities), collectively known as homology classes [178]. These classes capture fundamental topological features that are robust to deformation and noise. The premier tool for analyzing these features in the context of noisy, real-world data, such as neural signals, is Persistent Homology (PH). By analyzing a simplicial complex across a filtration of connectivity thresholds, PH robustly identifies topological features that persist across a wide range of scales, distinguishing them from transient, noise-induced artifacts [185], [186]. The output is typically visualized as a persistence diagram or barcode, which plots the birth and death of each topological feature, with its lifetime indicating its structural significance These topological fingerprints offer a powerful lens through which to characterize the intrinsic shape of complex networks, revealing organizational principles beyond pairwise interactions [152], [154].

This approach has yielded profound insights into brain architecture. In a landmark study, Sizemore et al. (2018) employed persistent homology to identify and characterize topological cavities in the structural connectome, revealing a scaffold of cyclical pathways that are hypothesized to support functional segregation and distributed processing [153]. Underscoring the power of this approach, Billings et al. (2021) conducted a direct comparison and found that metrics derived from persistent homology (specifically, distances between persistence diagrams of Betti numbers $H_0$, $H_1$, and $H_2$) significantly outperformed combinatorial and standard graph metrics in differentiating dynamic brain states during cognitive tasks [179]. Furthermore, this framework has been used to critically examine the very existence of higher-order structures in functional brain networks, highlighting the statistical challenges and the necessity of rigorous correction for multiple comparisons when inferring high-dimensional simplices from correlation data [180].

The current frontier in this domain elevates the simplicial complex from a static description to a dynamic scaffold for analyzing neural activity, a field known as Topological Signal Processing (TSP). This paradigm treats neural signals as simplicial signals defined on the nodes, edges, and higher-dimensional simplices of the complex. A major thrust within TSP is the development of methods to construct instantaneous, time-varying Simplicial Complexes. One such approach is based on co-fluctuation, where the element-wise product of z-scored time series is used to define the weight of simplices at each moment in time [187], [188]. Another innovative technique, Multiplication of Temporal Derivatives (MTD), uses the product of the signals' temporal derivatives to capture high-resolution co-fluctuation dynamics [189], [190]. Once these dynamic Simplicial Complexes are constructed, their evolution can be analyzed using Persistent Homology to extract higher-order topological features, such as quadruplet interaction signatures and time-varying voids [190], or to derive a homological scaffold that identifies the critical pairwise connections underpinning the brain's large-scale cyclical organization [187]. A different branch of TSP, exemplified by Bispo et al. (2025), leverages Hodge theory to decompose the signal flow on a pre-constructed simplicial complex. This allows for the computation of the simplicial divergence (identifying information sources/sinks) and curl (quantifying rotational information flow), providing a physically meaningful interpretation of information processing in the brain [126]. These advanced methods represent a significant step towards a unified understanding of how the brain's higher-order structure shapes its complex, moment-to-moment dynamics.

*2.2.2.3. Comparative Analysis and Concluding Remarks*
The exploration of hypergraphs and TDA reveals two powerful yet conceptually distinct paradigms for explicitly modeling the higher-order architecture of brain networks. While both frameworks transcend the limitations of pairwise graphs, they offer complementary insights rooted in their different mathematical foundations. Hypergraphs, with their flexible, set-based definition, excel at addressing questions of membership and community [156], [191].They are optimally suited for identifying specific, collaborating ensembles of brain regions—the functional committees—without imposing restrictive assumptions on their internal structure. This makes them particularly powerful for applications focused on discovering task-specific functional groups or identifying biomarkers based on the participation of nodes in particular higher-order circuits [147], [149]. In contrast, TDA and its use of Simplicial Complexes are designed to answer questions about the global shape and topology of neural data [178], [186]. By focusing on hierarchical structures and identifying features like cycles and cavities, TDA provides a multi-scale topographical map of the interaction landscape. This approach is less concerned with the exact membership of any single group and more with the overall organizational principles—the loops, voids, and robust topological features that govern information flow and functional segregation across the entire brain connectome [144], [153].

In conclusion, the development of frameworks for explicit higher-order modeling represents a significant and complementary frontier in network neuroscience. By moving beyond purely statistical measures of multi-way dependence, these methods—whether through the combinatorial flexibility of hypergraphs or the topological richness of Simplicial Complexes—provide a structural and often more intuitive language for describing the complex interactions that underpin brain function. As demonstrated throughout this part, these advanced models have shown considerable potential to uncover novel organizational patterns [153], enhance the accuracy of clinical classifications [147], [149], and offer unique insights into the brain's dynamic architecture [171], [179]. While they do not replace the valuable information provided by implicit or pairwise methods, they reveal a layer of structural organization that often remains inaccessible to those approaches. Despite ongoing challenges in construction, analysis, and neurobiological interpretation, the continued development and integration of these explicit modeling paradigms promise to significantly deepen our understanding of the intricate architecture of the human brain

## 2.3. Hybrid Methods: Encoding Explicit Motifs in Pairwise Structures

After introducing implicit and explicit methods in modeling higher-order connections in the brain and classifying articles into these two main groups, one study remains that, according to the authors, has characteristics of both of these categories. This method follows a two-step process: first, it explicitly identifies higher-order structural units within a base network, and second, it implicitly encodes the information about these units into the weights of a new, enriched pairwise graph. This approach combines the conceptual clarity of explicit structures with the analytical tractability of traditional graph-theoretic tools.

This hybrid philosophy is the motif-based network construction framework. This paradigm defines higher-order interactions as the presence of specific, recurring network motifs, which are considered fundamental units of information processing. The process begins with the construction of a base pairwise network, often directed to capture information flow (e.g., using Granger Causality). A statistical search is then conducted to explicitly identify over-represented motifs, such as specific three-node directed triplets.

The key innovation lies in the subsequent encoding step. A new motif-based adjacency matrix is created where the weight of the edge between two nodes implicitly represents their co-participation in the identified significant motifs. The resulting network is therefore hybrid in nature: while its structure is pairwise (an N×N matrix), its content is fundamentally higher-order, as each edge weight embeds information about explicit triplet interactions. This hybrid-order network, which uniquely captures local, directed circuit patterns, can then be analyzed using advanced tools like Graph Neural Networks (GNNs), demonstrating the power of integrating explicit higher-order information within a computationally familiar pairwise framework [192].

Another hybrid approach first constructs an explicit data-driven hypergraph, for instance using k-nearest neighbors to define hyper edges, and then leverages this high-order structural information to enrich the estimation of a dynamic functional connectivity network. Teng et al. (2024) demonstrated this by combining a hypergraph manifold regularizer with a model of directed temporal FC, resulting in a final pairwise network that implicitly encodes higher-order neighborhood information [175].

## 2.4. Higher-Order Analysis

While the majority of this review has focused on the rich and dynamic landscape of higher-order functional interactions, a complete picture of brain organization requires consideration of the underlying anatomical scaffold upon which these dynamics unfold. The study of higher-order structural interactions aims to characterize complex organizational principles within the brain's physical architecture. These methods, though less numerous than their functional counterparts, may provide crucial insights into the static blueprint that constrains and shapes brain function. The literature in this domain has primarily evolved along two distinct lines, distinguished by the imaging modality used: those based on diffusion imaging to map the brain's white matter wiring [193], and those based on T1-weighted imaging to analyze gray matter morphology [194], [195], [196].

One prominent approach for inferring higher-order structural relationships leverages data from Diffusion Spectrum Imaging (DSI) to model the brain's white matter connectome. Going beyond direct tractography, these methods seek to capture the broader topological context of each brain region. Inspired by natural language processing, the connectome embedding framework uses random walk-based algorithms like node2vec to traverse the structural graph. This process generates a low-dimensional vector representation, or embedding, for each brain region. Crucially, these embeddings implicitly encode higher-order information by capturing multi-hop neighborhood relationships and community structure. The resulting vector space allows for the powerful prediction of functional connectivity from the structural backbone and the simulation of network-wide effects of localized lesions, revealing topological properties inaccessible through direct pairwise analysis alone [193].

A parallel line of inquiry derives higher-order information from morphological features extracted from T1-weighted MRI scans. In this paradigm, a morphological similarity network is first constructed, where the connection strength between two brain regions is defined by the similarity of their anatomical characteristics (e.g., cortical thickness, volume, or surface area). To extract higher-order relationships from this base network, a strategy conceptually analogous to topological HOFC is employed. The morphological profile of each region—its pattern of similarity to all other regions—is computed, and the correlation between these profiles yields a higher-order morphological network [194], [195]. This framework has been used to develop reliable, individualized biomarkers for brain age estimation [195] and has been further refined by combining it with techniques like the Minimum Spanning Tree (MST) to enhance the classification of major depressive disorder [196]. Collectively, these methods demonstrate that the covariance of morphological properties across the cortex reveals a hidden layer of structural organization with significant clinical relevance.

# 3. Challenges, Limitations, and Critical Perspectives

While the previous section detailed the remarkable progress and immense potential of higher-order methods, a comprehensive review requires a critical examination of their inherent challenges and limitations. The field of higher-order brain analysis is still in a formative stage, and its methods, though mathematically elegant, are accompanied by a host of conceptual, methodological, and practical hurdles. A nuanced understanding of these challenges is not only essential for the cautious interpretation of current findings but is also crucial for guiding the future development of more robust and neurobiologically grounded approaches. This section provides a critical perspective on the field, organized around four central themes: conceptual and interpretive challenges, methodological and statistical pitfalls, computational and scalability issues, and finally, data and generalizability concerns.

## 3.1. Conceptual and Interpretive Challenges

In higher-order neuroscience, many of the open challenges are conceptual rather than purely computational. A recurring question concerns what these complex mathematical constructs actually represent in the biological context of the brain. Bridging the gap between mathematical abstraction and neurobiological interpretation remains a key difficulty, as discussed in recent theoretical works. This tension is reflected in the ambiguity of what higher-order measures capture and in the challenge of distinguishing genuine higher-order interactions from effects that may arise from lower-order statistical dependencies.

A primary and pervasive challenge is the biological interpretation gap. This issue manifests in two distinct facets. Many higher-order representations are abstractions whose direct physiological correlates are not immediately obvious. For instance, the precise neurobiological meaning of a COC framework remains elusive [15], [27], [29], and it has been noted that these features lack specific neuronal correlates at the group level [27]. Likewise, while a topological cavity in TDA is a well-defined mathematical object, its exact neurobiological role is still a subject of active investigation [153]. Similar interpretive challenges arise in temporal higher-order and edge-centric approaches, where linking a feature back to individual brain regions is often intractable [85], complicating its interpretation [81], [90].

The second facet of this gap is the lack of neurobiological generative models. The majority of current methods are descriptive and data-driven; they excel at identifying that a statistical pattern exists but offer little insight into why it exists, i.e., the underlying neural mechanisms [6]. This is a noted limitation in hypergraph and dynamic HOFC models constructed via non-biological criteria like KNN [11], [160] and in CoC-based methods that are purely data-driven [29]. Without clear generative models, it remains difficult to move from observing higher-order patterns to truly understanding their causal role in brain function.

A closely related issue is the genuineness problem: determining whether an observed higher-order statistical dependency represents a truly irreducible, synergistic interaction or is merely a manifestation of complex, lower-order (pairwise) effects. Several critical studies have demonstrated that some higher-order phenomena can be surprisingly well-replicated by null models based on static, pairwise covariance structures alone, questioning whether they provide genuinely new information [197], [198].This has been most notably argued in the context of the edge-centric paradigm, where high-amplitude events and the broader eFC matrix were shown to be largely predictable from the static functional connectome [197]. This conflation of true higher-order synergy with the coincidental accumulation of pairwise effects is a general concern across many frameworks [198]. It has motivated the development of methods specifically designed to isolate genuine interactions, such as those based on multivariate cumulants [101] and information-theoretic decomposition (e.g., O-information and PID) [131], [138], though even these advanced tools face challenges, such as the potential for O-information to conflate synergy and redundancy in small systems [138].

Furthermore, the increasing sophistication of higher-order methods often introduces a black box problem, sacrificing interpretability for predictive power. The use of clustering to reduce the dimensionality of dHOFC networks, for example, makes it difficult to trace the final predictive features back to the original brain connections, leading to a loss of neurobiological interpretability [11]. This challenge is even more pronounced in end-to-end deep learning models. While architectures that fuse multiple views of connectivity can achieve high classification accuracy, their internal decision-making processes are often opaque, making it difficult to extract clear, falsifiable biomarkers from the learned features [51].

Finally, some foundational approaches lack a solid theoretical underpinning. The intuitive CoC concept, for example, is a heuristic that has been criticized for not being grounded in a rigorous theoretical framework [197]. Similarly, a critical perspective on the edge-centric paradigm suggests that many of its dynamic features may not fully exploit the temporal structure of the data and can be reproduced by static null models, questioning the theoretical added value beyond pairwise statistics [197]. This highlights an ongoing need to bridge the gap between heuristic innovations and established theoretical principles from fields like information theory and statistical physics, a direction pursued by frameworks such as O-information and multivariate cumulants [101], [114].

## 3.2. Methodological and Statistical Challenges

Beyond the fundamental challenges of interpretation, the practical application of higher-order methods is fraught with significant methodological and statistical hurdles. These issues pertain to the validity, reliability, and reproducibility of the results, questioning not just what the measures mean, but whether they can be robustly and accurately estimated from complex neuroimaging data. These challenges can be broadly grouped into three categories: sensitivity to analytical choices, concerns about statistical validity, and the heavy burden of underlying model assumptions.

A primary threat to the reproducibility of higher-order research is the sensitivity of results to the parameters and heuristic choices. Many dynamic methods, for instance, rely on a sliding-window approach, where the choice of window length and step size is often a dilemma with no gold standard; shorter windows capture rapid fluctuations but are susceptible to noise, while longer windows provide more stable estimates at the cost of temporal precision [13], [55], [57]. Similarly, clustering-based methods like dHOFC are highly dependent on the chosen number of clusters, a hyper parameter that can severely influence the discriminative ability of the resulting network [14]. This reliance on arbitrary or data-dependent parameters extends to explicit methods as well, from the selection of thresholds for constructing Simplicial Complexes [110] to the choice of a specific redundancy function in Partial Information Decomposition, where different choices can lead to strikingly different results [138]. As a whole, this lack of standardized pipelines complicates the comparison of findings across studies and raises concerns about the generalizability of results.

Another statistical challenge lies in ensuring the validity of inferences and mitigating the risk of spurious findings. The hierarchical nature of many higher-order methods, such as CoC, can lead to the accumulation of statistical error at each step of the calculation, potentially amplifying noise from the initial low-order estimates [25], [107]. Moreover, the vast search space inherent in higher-order analysis creates a severe multiple comparisons problem, increasing the likelihood of detecting spurious patterns that arise from chance alone, especially with limited data [134]. Some work has questioned the statistical validity of some dynamic higher-order phenomena altogether, demonstrating that key features of the edge-centric paradigm, such as high-amplitude events, can be largely reproduced by static, Gaussian null models that lack any temporal dynamics [77], [197]. This highlights a crucial need for appropriate null models and rigorous statistical testing to distinguish genuine higher-order dynamics from the complex statistical footprint of lower-order structures [101].

Finally, many higher-order methods carry a burden of model assumptions that may not always hold true for noisy and complex neural data. Information-theoretic measures, for example, often rely on the assumption of Gaussianity for tractable estimation, an assumption that is frequently violated by real neural signals, potentially leading to inaccurate estimates of synergy and redundancy [116], [132]. The simplicial complex approach imposes strong structural assumptions; it require that all sub-faces of a simplex must also exist (the clique assumption), a constraint that may not be representative of all biological interactions [156], [199]. Furthermore, methods that rely on the stationarity of time series can be biased by non-stationary noise or task-related signal changes [52]. Even the fundamental properties of fMRI data, such as strong auto-correlation, pose significant challenges for statistical inference frameworks like bootstrapping and permutation testing, complicating the assessment of statistical significance for any higher-order measure [101]. These underlying assumptions must be carefully considered when applying and interpreting the results of higher-order analyses.

## 3.3 Computational and Scalability Challenges

Perhaps the most significant practical barrier to the widespread adoption of higher-order methods is the profound challenge of computational and scalability issues, rooted in the combinatorial complexity and the curse of dimensionality. As analytical approaches move from pairwise (order 2) to triplet (order 3), quadruplet (order 4), and beyond, the number of potential interactions to consider grows explosively. For a network of N nodes, the number of possible k-node interactions scales as $O(N^k)$, a combinatorial explosion that quickly renders exhaustive searches computationally infeasible even for moderately sized brain networks [6], [13], [108], [125], [131], [188], [197], [200], [201]. This fundamental challenge manifests across nearly all higher-order paradigms, imposing severe constraints on the scale and scope of analyses.

In implicit methods, this challenge is starkly evident. The original formulation of temporal HOFC, which involves correlating all pairs of ETS, results in a feature space that scales with $O(N^4)$, a complexity that is computationally prohibitive and requires vast amounts of memory [14], [197]. This has necessitated the widespread use of dimensionality reduction techniques, such as clustering ETS, as a standard but heuristic step in the pipeline [11], [13]. Information-theoretic measures face a similar barrier; the number of information atoms in frameworks like Partial Information Decomposition (PID) grows hyper-exponentially, making the analysis of systems with more than a handful of elements practically impossible [131], [138]. Even for more tractable measures like O-information, many studies are forced to limit their analysis to small brain parcellations or low orders (e.g., triplets) to manage the computational load [104].

Some explicit methods are equally, susceptible to these scalability issues. In TDA, the number of possible simplices in a complex can grow exponentially with the number of nodes, and the complexity of standard persistent homology algorithms can be cubic in the number of simplices [152], [202]. The task of simply counting k-cliques, a foundational step in many TDA and combinatorial approaches, is itself an NP-complete problem, meaning its computational cost increases exponentially with network density [184]. These computational demands often lead to practical limitations, such as memory exhaustion in software libraries [202] and forcing analyses to be restricted to lower sparsity levels or lower-order interactions [184], [187].

While various strategies have been proposed to mitigate these issues, such as greedy search algorithms in information theory [6], hierarchical partitioning in MVND [109], or preprocessing to collapse the size of a complex in TDA [152], computational scalability remains a central and active area of research, currently limiting the application of many higher-order methods to whole-brain, high-resolution connectomes.

## 3.4. Data and Generalizability Challenges

Beyond the conceptual and computational hurdles, the application of higher-order methods in neuroscience is fundamentally constrained by the nature of the available data and the overarching challenge of generalizability. These practical issues directly impact the reliability of estimations and the ultimate translational potential of the findings.

A core issue is the intrinsic hunger for data that characterizes most higher-order analytical techniques. The robust estimation of multivariate dependencies requires substantially more data than pairwise measures to achieve similar statistical power. This is a particularly acute problem for information-theoretic methods, where estimating joint probability distributions from a limited number of samples is often infeasible in practical scenarios [116], [138]. Similarly, dynamic methods like dHOFC, which rely on short, sliding time windows, operate in a low-sample regime that can lead to unreliable estimates and an increased risk of overfitting [14], [107]. This data hunger is often at odds with the inherent limitations of neuroimaging modalities like fMRI, which provide relatively short and noisy time series, creating a persistent tension between methodological sophistication and data availability.

This tension is compounded by challenges related to sample size, heterogeneity, and the generalizability of findings. Many pioneering studies in this emerging field have relied on relatively small or single-site datasets, which raises questions about the reproducibility and generalizability of their conclusions to broader populations [188]. Even when larger, multi-site datasets are used, site-specific heterogeneity can introduce significant confounds that are difficult to disentangle from true biological effects. Furthermore, the inherent noisiness of neural data means that higher-order metrics, which often involve multi-step calculations, can be particularly susceptible to noise propagation. For example, the reliability of dHOFC measures has been shown to be poor for weaker connections, which are more likely to be affected by noise[13]. Similarly, the quality of higher-order networks derived from MVND or topological HOFC is directly dependent on the reliability of the initial low-order FC estimates, which are themselves subject to noise and statistical uncertainty [37], [107]. Addressing these challenges through the development of more robust estimation techniques and validation across large, diverse datasets is a critical step for moving the field of higher-order connectomics from exploratory analysis toward robust scientific and clinical application.

### 3.5. Synthesis of Challenges and a Comparative Overview

In summary, the journey into higher-order brain analysis, while promising, is navigated through a landscape of significant challenges. These hurdles are not uniform across the field; different methodological paradigms are susceptible to different pitfalls to varying degrees. Table [Number] provides a comparative overview, summarizing the relative severity of the key conceptual, methodological, computational, and data-related challenges for each major class of methods discussed in this review.

Conceptually, the gap between mathematical abstraction and biological reality remains a universal concern, though it is perhaps most acute for highly abstract frameworks like TDA and heuristic approaches like CoC. Methodologically, methods relying on arbitrary parameter choices, such as the window length in dynamic analyses or the number of clusters in dHOFC, pose a greater threat to reproducibility. Computationally, the curse of dimensionality is a formidable barrier for nearly all paradigms, but it manifests most severely in methods that require exhaustive combinatorial searches, such as information-theoretic measures and TDA. Finally, the need for large, high-quality datasets is a shared vulnerability, especially for data-hungry approaches like deep learning and information theory. Acknowledging this complex tapestry of strengths and weaknesses is the first step toward developing more robust, interpretable, and ultimately more useful models of the higher-order brain.

Table 1. A Comparative Overview of Challenges Across Major Higher-Order Analysis Paradigms. (✓ = Moderate Challenge; ✓✓ = Significant Challenge; ✓✓✓ = Severe Challenge; - = Not a primary challenge / Addressed by design)

| Challenge Category | Challenge | Topological HOFC | Temporal HOFC | Multivariate Statistics | Information Theory | Hypergraphs | Simplicial Complexes (TDA) |
|---|---|---|---|---|---|---|---|
| Conceptual & Interpretive | Biological Interpretation Gap | ✓✓✓ | ✓✓✓ | ✓✓ | ✓✓ | ✓✓ | ✓✓✓ |
| Conceptual & Interpretive | "Genuineness" Problem | ✓✓✓ | ✓✓✓ | - | - | ✓✓ | ✓ |
| Conceptual & Interpretive | Black Box Problem | ✓ | ✓✓ | ✓ | ✓ | ✓✓ | ✓ |
| Methodological & Statistical | Sensitivity to Parameters | ✓ | ✓✓✓ | ✓✓ | ✓✓ | ✓✓ | ✓✓✓ |
| Methodological & Statistical | Statistical Validity & False Positives | ✓✓ | ✓✓ | ✓✓ | ✓✓✓ | ✓✓ | ✓✓✓ |
| Methodological & Statistical | Burden of Model Assumptions | ✓ | ✓ | ✓✓✓ | ✓✓✓ | ✓ | ✓✓✓ |
| Computational & Scalability | Combinatorial Complexity | - | ✓✓ | ✓✓ | ✓✓✓ | ✓✓✓ | ✓✓✓ |
| Data & Generalizability | Hunger for Data | ✓ | ✓✓ | ✓✓ | ✓✓✓ | ✓✓ | ✓✓ |

## 4. Discussion, Future Directions, and Conclusion

Our survey has traversed the vast and rapidly evolving landscape of higher-order interactions in brain network neuroscience. We have tried to chart the major methodological currents, categorizing them into two principal paradigms: implicit approaches that quantify the statistical signature of group dependencies, and explicit approaches that construct tangible mathematical objects like hypergraphs and Simplicial Complexes. This work has tried to reveal a rich and diverse toolkit for moving beyond the limitations of traditional pairwise analysis. Yet, it has also highlighted that the pursuit of higher-order understanding is not a linear progression toward ever-increasing complexity, but a nuanced exploration filled with conceptual challenges, methodological trade-offs, and critical open questions. This final part aims to synthesize these findings, critically discuss the overarching challenges facing the field, and outline promising directions for future research.

## 4.1. Synthesis and Comparative Insights

**Implicit vs. Explicit: A Symbiotic Relationship, Not a Dichotomy:** A primary insight emerging from this review is that the implicit and explicit paradigms, while distinct in their methodological approaches, are not adversaries but rather exist in a symbiotic and increasingly convergent relationship. The choice between quantifying the statistical strength of an interaction (the forte of implicit methods) versus describing its geometric structure (the forte of explicit methods) is less a technical dichotomy and more a philosophical decision dictated by the research question. Crucially, this distinction is blurring. As we have seen, the lines between paradigms merge as methods from one inform the other: information-theoretic metrics, developed to implicitly quantify dependency, are now being used as principled tools to guide the construction of explicit hypergraphs and simplicial complexes [125], [126]. Conversely, tools from algebraic topology can be used to extract implicit, scalar measures of network organization [184]. This symbiosis underscores a fundamental truth: the choice of representation shapes the achievable insights, with different formalisms capable of revealing entirely different organizational principles—such as core-periphery structures, topological holes, or community organization—from the very same underlying system [199].

**The Complexity vs. Utility Trade-off: Is More Always Better?** A second critical insight is that greater methodological complexity does not guarantee superior performance in practical applications. While higher-order models offer greater theoretical richness, their utility is context-dependent. Several studies report that simpler, lower-order models can achieve comparable or even better performance in certain clinical classification tasks. For instance, a study differentiating schizophrenia from bipolar disorder found that static functional connectivity provided greater discriminative power than dHOFC [16]. This observation is further supported by findings that the incremental benefit of adding higher-order information may diminish as the order of interaction becomes very large [203], and that for some cognitive tasks, traditional pairwise FC remains among the most effective decoding methods [114]. This complexity-utility trade-off can be attributed to several factors discussed in the previous chapter, including the high variance of higher-order estimators, the risk of overfitting in high-dimensional feature spaces, and the possibility that the most robust biomarkers for a given condition are embedded in stable, lower-order connectivity patterns.

**A Common Ground: Convergent Applications in Neuroscience:** Despite their methodological diversity, a striking convergence is evident in the application of higher-order methods to clinical and translational neuroscience. Across both implicit and explicit paradigms, these advanced models have consistently demonstrated an enhanced sensitivity for biomarker discovery and disease classification. Frameworks ranging from CoC-based HOFC [11], [12] and edge-centric networks to information-theoretic measures [6] and hypergraphs [71], [163] have all been successfully employed to improve the diagnosis of a wide spectrum of neurological and psychiatric conditions, including Alzheimer's disease, schizophrenia, autism, and depression. Beyond classification, these methods are also showing promise in predicting clinical outcomes and behavioral traits, offering new tools for assessing post-stroke recovery [23] or individual differences in cognitive traits like risk propensity [87]. This consistent success across different diseases and methodologies underscores a fundamental point: pathology sometimes manifests in the disruption of complex, multi-way interactions that are invisible to traditional pairwise analysis.

**Unveiling Fundamental Principles of Brain Organization:** In parallel, higher-order methods are providing profound new insights into the fundamental principles of brain organization and cognitive function. Explicit modeling paradigms like hypergraphs and Simplicial Complexes are being used to map the scaffold of the connectome, revealing previously uncharacterized features such as higher-order hubs [125] and topological cavities [153]. Dynamic approaches, particularly the edge-centric paradigm, are revolutionizing our understanding of fine-scale temporal organization by linking transient co-activation events to cognitive states and underlying structural modules [75], [78]. Meanwhile, information-theoretic frameworks are beginning to quantify the intricate balance between information integration and segregation that supports complex cognition [7], [124]. Collectively, these applications highlight that higher-order analysis is not merely a more complex tool, but an essential one for addressing foundational questions about how the brain's architecture gives rise to its rich functional and cognitive repertoire.

## 4.2 Grand Challenges and Future Directions

The maturation of higher-order neuroscience depends on addressing several grand challenges that span the conceptual to the practical. A primary challenge remains the gap between statistical observation and biological mechanism. Many current methods are descriptive, identifying that a higher-order pattern exists without explaining how it is generated by the underlying neural circuitry. The future of the field lies in moving from purely data-driven models to neurobiologically plausible generative models [204]. Such models, which could incorporate principles like oscillatory dynamics, synaptic plasticity, and neuromodulation, are needed to test hypotheses about how emergent, higher-order phenomena arise from local circuit rules [10], [78].

Brain organization is inherently multi-modal and multi-scale. Future research must move towards integrating information across imaging modalities (e.g., fMRI, EEG/MEG, DTI) and spatial scales. Frameworks that can harmonize different types of relationships, such as functional co-fluctuations and structural pathways, within a unified higher-order model, like the edge covariance approach [98], represent a promising direction. This integration is critical for understanding the complex interplay between the brain's static structural scaffold and its dynamic functional repertoire.

The proliferation of novel methods has created a fragmented research landscape [114] that hinders reproducibility and widespread adoption. Unlike pairwise analysis, which is supported by numerous standardized toolboxes, many higher-order methods lack accessible, well-documented implementations. The development of comprehensive, open-source toolboxes that implement multiple higher-order techniques is a critical need. While excellent resources for specific paradigms are emerging [Ref: e.g., HOC-Toolbox for CoC; Ripser, Gudhi for TDA - Otter et al., 2017], a unified platform would significantly accelerate progress and improve the comparability of studies.

As highlighted by several critical studies, the risk of inferring spurious higher-order structure from noise or from the complex footprint of lower-order correlations is substantial [101], [197]. A major future direction must be the development and consistent application of appropriate null models and rigorous statistical inference frameworks. This includes moving beyond simplistic null models that ignore the temporal auto-correlation of neuroimaging data and developing principled methods to address the severe multiple comparisons problem inherent in any combinatorial search for higher-order patterns.

While many studies have demonstrated the potential of higher-order biomarkers, a significant gap remains between exploratory research and robust clinical translation. Future work must prioritize validation on large, diverse, and multi-site clinical datasets to ensure that identified biomarkers are generalizable and not specific to a small, homogeneous sample. Furthermore, assessing the test-retest reliability of these complex metrics is a crucial, yet often overlooked, step for establishing their suitability as stable, longitudinal markers of disease progression or treatment response [13].

## 4.3 Concluding Remarks

The study of higher-order interactions is fundamentally reshaping our understanding of the brain, moving us from a network of simple pairwise connections to a vision of an entangled brain defined by complex, emergent, and multi-way relationships [10]. This review has charted the diverse and innovative methodologies that form the vanguard of this new frontier. From the implicit quantification of statistical dependencies to the explicit construction of topological structures, these tools provide a rich vocabulary for describing the intricate architecture of brain function. However, as we have discussed, these powerful methods are not a panacea. Their application demands a critical awareness of their conceptual ambiguities, statistical pitfalls, and practical limitations. The path forward lies not in a blind pursuit of complexity, but in a thoughtful integration of these novel approaches with rigorous neurobiological grounding. By embracing this challenge, the field of higher-order connectomics is poised to unlock profound new insights into the nature of cognition, health, and disease.

## AI Declaration

The authors acknowledge the use of artificial intelligence (AI) in the preparation of this manuscript. Specifically, the Gemini 2.5 Flash large language model was utilized to assist in summarizing a portion of the referenced literature and extracting key methodological information. Additionally, the Gemini 2.5 Pro model was employed as a writing assistant to help structure and articulate the concepts outlined in the authors' pre-existing mind map and detailed notes. The authors meticulously reviewed, edited, and revised all AI-generated text to ensure accuracy, originality, and alignment with the scientific narrative of the review. Full responsibility for the final content, including all arguments, interpretations, and conclusions, rests solely with the human authors.

## Supplementary Materials

The following supporting information can be downloaded with this article: Figure S1: A comprehensive mind map illustrating the structure, key concepts, and classification of methodologies discussed in this review.